\documentclass[sigconf]{acmart}

\AtBeginDocument{%
  
    }

\copyrightyear{2026}
\acmYear{2026}
\setcopyright{cc}
\setcctype{by}
\acmConference[CIKM '26]{Proceedings of the 35th ACM International Conference on Information and Knowledge Management}{November 07--11, 2026}{Rome, Italy}
\acmBooktitle{Proceedings of the 35th ACM International Conference on Information and Knowledge Management (CIKM '26), November 07--11, 2026, Rome, Italy}
\acmDOI{10.1145/3799682.3840157}
\acmISBN{979-8-4007-2539-5/2026/11}

\usepackage{graphicx}
\usepackage{xspace}
\usepackage{multirow}
\usepackage{enumitem}

\newcommand{\method}{\textsc{ConnectionMind}\xspace }

\begin{document}

\title[ConnectionMind]{ConnectionMind: Leveraging Social Networks and Large Language Models for Personalized Recommendation at Meta}

\author{Haoyu Han}
\affiliation{%
  \institution{Michigan State University}
  \city{East Lansing}
  \state{Michigan}
  \country{USA}
}
\email{hanhaoy1@msu.edu}

\author{Yuming Liu}
\affiliation{%
  \institution{Meta Platforms, Inc.}
  \city{Menlo Park}
  \state{California}
  \country{USA}}
\email{yumliu@meta.com}

\author{Lei Huang}
\affiliation{%
  \institution{Meta Platforms, Inc.}
  \city{Menlo Park}
  \state{California}
  \country{USA}}
\email{leihuang@meta.com}

\author{Lizhu Zhang}
\affiliation{%
  \institution{Meta Platforms, Inc.}
  \city{Menlo Park}
  \state{California}
  \country{USA}}
\email{lizhu@meta.com}

\author{Jiliang Tang}
\affiliation{%
  \institution{Michigan State University}
  \city{East Lansing}
  \state{Michigan}
  \country{USA}
}
\email{tangjili@msu.edu}

\author{Xiangjun Fan}
\affiliation{%
  \institution{Meta Platforms, Inc.}
  \city{Menlo Park}
  \state{California}
  \country{USA}}
\email{maxfan@meta.com}

% \affiliation{%

% \affiliation{%

% \affiliation{%

% \affiliation{%

% \affiliation{%

% \affiliation{%

\renewcommand{\shortauthors}{Haoyu Han et al.}

\begin{abstract}
Modern recommendation systems on social media platforms such as Meta must model complex social relationships alongside massive and heterogeneous content such as text and video. Traditional recommendation models, however, often omit these signals or treat them independently, limiting their ability to integrate multi-relational context for fine-grained personalization. We present \method, a production-ready recommendation framework that tightly integrates social network structure with large language models (LLMs) to enable scalable and reasoning-aware personalization at Meta. \method constructs a heterogeneous graph connecting users, items, friends, groups, and creator pages, and formulates recommendation as a graph reasoning problem that discovers personalized paths from users to candidate items. An LLM-based policy reasons over these graph structures to guide recommendation decisions. To train the system at scale, \method adopts a two-stage strategy: supervised fine-tuning (SFT) on large-scale user--item interaction trajectories to initialize the reasoning policy, followed by end-to-end reinforcement learning (RL) to further optimize personalized graph reasoning. Extensive experiments on multiple real-world datasets demonstrate consistent improvements over representative baselines. More importantly, \method has been deployed in Meta's large-scale recommendation pipeline and evaluated through online A/B tests, achieving a \textbf{0.43\% improvement in video watch time}. These results demonstrate the effectiveness and measurable real-world impact of graph-based LLM reasoning for production recommendation.

\end{abstract}



% %% A "teaser" image appears between the author and affiliation
% %% information and the body of the document, and typically spans the


\maketitle

% \vspace{-0.1in}
\section{Introduction}
Recommendation systems~\cite{ko2022survey, li2024recent, han2020stgcn} are the backbone of modern online platforms, enabling users to efficiently discover relevant content from an overwhelming volume of information. From e-commerce to short-video and social media platforms, these systems connect users with products, videos, and posts that align with their interests and engagement behaviors. Traditional approaches~\cite{koren2009matrix, he2017neural, cheng2016wide}, such as collaborative filtering and deep neural recommendation models, primarily rely on user–item interaction histories to learn personalized preferences. More recently, recommender systems~\cite{wu2024survey, liu2024rec} have incorporated rich side information, including text, images, and videos, and benefited from large-scale foundation models such as large language models (LLMs)~\cite{zhao2023survey, touvron2023llama} and large vision models (LVMs)~\cite{liu2024sora, zhang2024vision}, which enable high-level semantic understanding of multimodal content beyond simple interaction patterns.

However, understanding users in isolation is insufficient for social platforms~\cite{fan2019graph, tang2016recommendation}, where engagement is strongly shaped by social relationships and community dynamics. On social platforms such as Meta, users often interact with content that is endorsed by friends with shared interests, originates from creators they follow, or is discussed within groups they belong to. Friend networks, group memberships, and creator interactions encode signals of trust, influence, and shared preferences that cannot be captured by user–item interactions alone.

To leverage social signals, a broad spectrum of social recommendation methods~\cite{tang2013social, li2023survey} augments recommender systems with social networks, ranging from social regularization~\cite{ma2008sorec} and trust-aware factorization~\cite{guo2015trustsvd} to diffusion-based models~\cite{wu2019neural, li2024recdiff} and graph neural networks (GNNs)~\cite{fan2019graph}. Despite their differences, most existing approaches share a common principle: social information is propagated or aggregated across the network and compressed into latent user or item representations for downstream recommendation. Although effective, this paradigm faces several challenges in large-scale social recommendation.
First, many methods~\cite{wu2019neural, wu2020diffnet++} rely on broad propagation over social neighborhoods under fixed structural assumptions, while only a small, context-dependent subset of social relations may be relevant to a given recommendation. Indiscriminate aggregation can therefore introduce substantial noise. Second, collapsing heterogeneous social evidence into dense embeddings limits transparency into which signals drive a recommendation, creating challenges for debugging, monitoring, and trust in production systems and motivating recent work on explainable social recommendation~\cite{guo2025sorex}. Third, most social recommenders treat semantic information, such as creator attributes or content descriptions, as static features rather than context-dependent signals that can guide which relations to follow and which evidence to trust. These limitations reveal a fundamental gap in current social recommendation paradigms: the lack of a mechanism for selectively reasoning over social structure in a decision-driven manner.

To address these limitations, we model the platform as a heterogeneous social–item interaction graph that connects users, friends, groups, creator pages, and items. Users are linked to their friends, group memberships, followed pages, and consumed items, while items connect to engaged users, their creator pages, and related content. This representation makes social context explicit and provides a structured substrate for modeling how preferences and influence propagate through the platform. Building on this graph, we propose \method, a production framework at Meta that integrates social graph modeling with a learned reasoning policy for personalized recommendation. \method casts recommendation as a \textbf{path-based graph exploration problem}: given a user node, the policy explores a compact neighborhood on the graph to identify high-value paths that lead to candidate items, naturally supporting socially grounded personalization and interpretable evidence traces.

Designing such a system introduces several practical challenges. The reasoning policy must make structured decisions over a heterogeneous graph, i.e., generating relation-consistent actions and coherent paths rather than free-form text. Moreover, production constraints require the policy to operate with compact, task-relevant context, since the full social–content graph is prohibitively large to be included as the model's input. To address these challenges, \method employs a task-aware subgraph sampling strategy to construct compact neighborhoods, and adopts a two-stage training pipeline: (i) supervised fine-tuning (SFT) on large-scale interaction trajectories to warm-start the policy to produce valid, preference-aligned reasoning paths, followed by (ii) end-to-end reinforcement learning (RL) that further improves exploration and recommendation decisions by optimizing task rewards. Our key contributions are summarized as follows:
% \vspace{-0.1in}
\begin{itemize}[leftmargin=1em]
    \item We introduce \method, a production-ready framework that reformulates social recommendation as query-conditioned path exploration over a heterogeneous social--item graph.
    \item We design an LLM-guided reasoning policy, trained with SFT followed by end-to-end RL, to selectively discover informative social and semantic paths for recommendation.
    \item We demonstrate the effectiveness and practicality of \method through public benchmark experiments and online A/B testing in Meta's production system, showing measurable gains across key engagement metrics.
\end{itemize}

% \vspace{-0.1in}
\section{Related Work}
% In this section, we briefly review the literature on social recommendation and LLM-based recommendation.
% \vspace{-0.1in}
\subsection{Social Recommendation}
Social recommendation extends conventional user--item frameworks by leveraging social relations such as friendship, trust, and user influence. Early methods incorporate social information through regularization or trust-aware constraints, encouraging socially connected users to share similar representations and alleviating sparsity and cold-start issues. Representative approaches include SoRec~\cite{ma2008sorec}, TrustSVD~\cite{guo2015trustsvd}, and  TrustWalker~\cite{jamali2009trustwalker}. These methods are simple and scalable, but typically capture social influence only indirectly.
For platforms with multiple entity and relation types, heterogeneous information networks (HINs) provide a natural abstraction. Meta-path-based methods, such as metapath2vec~\cite{dong2017metapath2vec} and HERec~\cite{shi2018heterogeneous}, leverage typed paths to encode heterogeneous semantics. With the development of graph neural networks (GNNs), social recommendation increasingly models coupled user--user and user--item graphs through message passing. GraphRec~\cite{fan2019graph} jointly aggregates social and interaction signals, while MHCN~\cite{yu2021self} introduces multi-channel hypergraph convolution with self-supervision to capture high-order social relations. SEPT~\cite{yu2021socially} further improves robustness through socially-aware self-supervised learning. Although effective, these methods generally compress social context into dense embeddings through neighborhood aggregation.

Another line of work models social influence as a diffusion process. DiffNet~\cite{wu2019neural} recursively propagates preferences along social links, and DiffNet++~\cite{wu2020diffnet++} extends this framework by jointly modeling influence and interest diffusion. Recent methods further address noisy social graphs: RecDiff~\cite{li2024recdiff} applies diffusion-based denoising in representation space, while SGIL~\cite{yang2025invariance} formulates social recommendation as invariant learning across noisy environments. Overall, prior social recommenders demonstrate the value of peer and community context, but most follow an aggregation or diffusion paradigm that offers limited selectivity and interpretability over which social evidence supports a recommendation.

% \vspace{-0.1in}
\subsection{LLM-based Recommendation}
Large language models (LLMs) introduce a new paradigm that frames recommendation as language understanding and generation. Early work reformulates recommendation tasks into text-to-text objectives, enabling a single model to support rating prediction, ranking, explanation, and sequential recommendation. P5~\cite{geng2022recommendation} and its extensions~\cite{xu2023openp5, zhang2025recommendation} exemplify this direction, while several studies use LLMs as ranking or re-ranking components. LLaMARec~\cite{yue2023llamarec} and RecRanker~\cite{luo2025recranker} re-rank candidate sets generated by conventional recommenders, balancing semantic reasoning with efficiency. Other methods explore generative recommendation, where LLMs directly generate item identifiers or item-representative tokens. TIGER~\cite{rajput2023recommender} and BIGRec~\cite{bao2023bigrec} adopt grounding strategies to bridge language generation and discrete item spaces. 
% Hybrid methods further inject collaborative signals into LLMs: CoLLM~\cite{zhang2025collm} integrates collaborative embeddings into the LLM token space, while A-LLMRec~\cite{kim2024large} bridges frozen CF models and frozen LLMs.

Our work lies at the intersection of social recommendation and LLM-based recommendation. 
Unlike prior LLM-based recommendation methods, \method learns a structured exploration policy over a heterogeneous social--item interaction graph. This design enables selective and interpretable use of social evidence while remaining compatible with large-scale recommendation pipelines.

% \vspace{-0.1in}
\section{Social Graph and Path Formulation}
\label{sec:graph}
We study large-scale personalized video recommendation in a social platform setting, where the goal is to surface engaging videos for a target user based on both individual preferences and social context. Unlike traditional recommendation scenarios that rely primarily on user--item interaction histories, social platforms such as Meta expose users to content through a rich ecosystem of social relations, including friendships, followed pages (creators), group memberships, and content sharing behaviors. These social signals encode trust, influence, and shared interests, and play a central role in shaping user engagement.

Through extensive analysis of production-scale interaction logs, we observe three consistent patterns in social video consumption at Meta. Due to privacy and platform policy constraints we report these findings qualitatively, and validate their impact via offline and online evaluations (Sections~\ref{sec:online}).
\textbf{First, social signals are essential for recommendation}: a substantial fraction of user engagements involve content that is connected to the user’s social context, such as items consumed by friends, posted by followed pages or creators, or shared within groups. 
\textbf{Second, social relevance is highly selective}: for a given user, engagement overlap is concentrated among a small subset of social entities, specific friends, followed pages, or groups, rather than being uniformly distributed across all connections. 
\textbf{Third, this selectivity is temporally stable}: users tend to repeatedly exhibit similar engagement overlap with the same or similar social entities over time, indicating persistent social influence rather than transient co-consumption effects. These observations motivate a model that can explicitly represent heterogeneous social structure and selectively identify compact, user-specific social evidence for recommendation.

% To capture the above characteristics, we model the platform as a heterogeneous social--item interaction graph that explicitly represents users, content, creators, and their diverse relations. We formulate recommendation as a path-finding problem on this graph, where relevant items are reached by traversing a compact set of informative social paths. This graph provides a structured substrate for our recommendation policy, enabling selective use of social signals and explicit, interpretable reasoning paths.

\vspace{-0.1in}
\subsection{Heterogeneous Social–item Interaction Graph Construction}

\begin{figure}[!htb]
    \centering
    \includegraphics[width=0.8\linewidth]{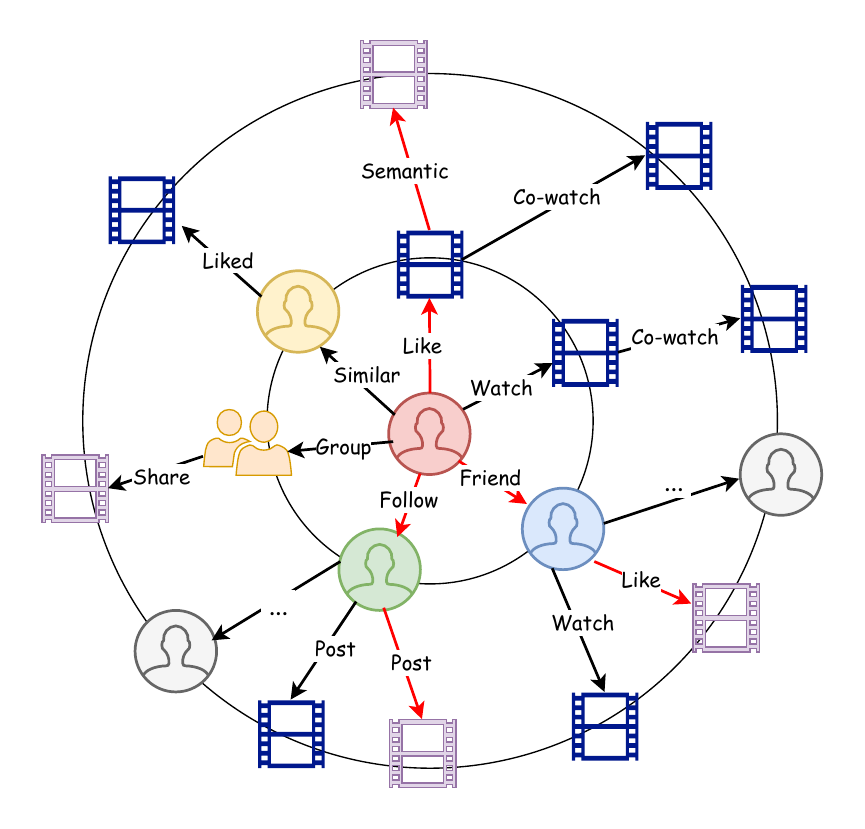}
    \vspace{-0.1in}
    \caption{An illustration of the heterogeneous social-item interaction graph}
    \label{fig:graph}
    \vspace{-0.2in}
\end{figure}

To support selective reasoning over social context, we construct a heterogeneous social--item interaction graph that explicitly encodes users, content, creators, and their diverse relations, as shown in Figure~\ref{fig:graph}. The graph serves as the structured search space for our recommendation policy: each typed edge represents a possible transition between social, creator, or item entities, and a sequence of such transitions can later be interpreted as evidence connecting a user to a candidate item. The graph is designed to (i) capture the multiple ways in which social signals influence video consumption, (ii) expose fine-grained relational structure for downstream path reasoning, and (iii) remain scalable under production constraints through task-driven sparsification.

\paragraph{Graph schema.}
We define a typed, temporal heterogeneous graph
\[
G_T = (V, E, \mathcal{R}), \quad V = \mathcal{U} \cup \mathcal{P} \cup \mathcal{I},
\]
where $\mathcal{U}$ denotes users, $\mathcal{P}$ indicates pages or creators, and $\mathcal{I}$ represents items (e.g., posts, videos, reels). Each edge $e \in E$ is associated with a relation type $r \in \mathcal{R}$, a timestamp $t_e$, where $T_{min} \leq t_e \leq T_{max}$, and a non-negative weight $w_e$ capturing interaction strength or semantic similarity.

\paragraph{Relation types.}
We instantiate multiple relation types to reflect distinct social and semantic mechanisms. All edges are time-stamped and directed unless otherwise specified.
\begin{itemize}[leftmargin=1em,itemsep=0pt,topsep=0pt]
    \item \textbf{User--User (Friend):} reciprocal edges represent explicit friendships, with weights encoding tie strength based on historical interactions.
    \item \textbf{User--User (Similar):} behaviorally similar users are connected based on engagement histories, excluding direct friends to distinguish homophily from explicit social ties.
    \item \textbf{User--Page (Follow):} edges indicate that a user follows a page or creator, with weights reflecting recency or engagement strength.
    \item \textbf{User--Group (Group):} edges connect users to groups they join or interact with.
    \item \textbf{Group--Item (Share):} edges connect groups to items shared or discussed within them.
    \item \textbf{Page--Item (Post):} each item is linked to the page or creator that published it.
    \item \textbf{Item--Item (Co-watch):} edges connect items frequently co-consumed by users within a temporal window, capturing behavioral co-engagement patterns.
    \item \textbf{Item--Item (Semantic Similarity):} edges connect items with high semantic similarity.
\end{itemize}

These relations jointly capture social influence, creator affinity, group context, collaborative filtering signals, and semantic relatedness. They also define different types of evidence routes for recommendation: a user may reach an item through a friend who engaged with it, a followed creator who posted it, a group where it was shared, or a previously consumed item that is behaviorally or semantically related to it. Thus, the graph provides both connectivity under sparsity and explicit relational structure for the path formulation introduced next.

\paragraph{Item representations (multimodal to text).}
To enable semantic reasoning over heterogeneous content, each item is converted into a compact textual representation. As our task focuses on video recommendation, for each video we extract on-screen text via OCR, apply ASR to audio tracks, and generate dense visual descriptions using a video large language model (VidLLM)~\cite{tang2025video}. All available signals are then concatenated and summarized by an LLM into a compact textual representation.

\paragraph{User and page representations.}
For users and pages, we derive node features by aggregating the textual summaries of their engaged or published items, respectively, together with available profile information. These aggregated signals are compressed via LLM-based summarization to produce compact semantic profiles that guide context-aware exploration.

\vspace{-0.1in}
\subsection{Recommendation as Path Discovery}
\label{subsec:sampling}

Given the heterogeneous social--item graph $G_T$, we formulate recommendation as discovering typed paths from a target user to candidate items. This formulation turns the graph constructed above into an actionable search space: instead of aggregating all neighboring information into a single representation, the model can selectively follow social, creator, group, and item relations that are relevant to the target user.

For a target user $u \in \mathcal{U}$, the goal is to identify items $i \in \mathcal{I}$ that the user is likely to engage with in the future. We define a reasoning path as
\[
\pi = (v_0{=}u, r_1, v_1, \ldots, r_L, v_L),
\]
where each hop $(v_{l-1}, r_l, v_l)$ corresponds to a valid typed edge in $G_T$, and the terminal node $v_L \in \mathcal{I}$ represents a candidate item. Each path provides an explicit chain of social or semantic evidence linking the user to the item.

Directly reasoning over the full graph is infeasible because each user may connect to many interacted items, friends, similar users, followed pages, and groups, and multi-hop expansion can quickly produce a large and noisy neighborhood. To address this challenge, we sample multiple compact, high-signal subgraphs for each target user. Each subgraph is centered at the user and preserves informative social and semantic context while remaining tractable. Specifically, for each relation type, we sample up to $M$ neighbors per node based on edge weights, such as interaction strength or recency, and semantic similarity derived from node representations. Repeating this process yields a set of $k$-hop local subgraphs, each capturing a complementary view of the user's social environment. These subgraphs provide compact inputs for downstream path reasoning while maintaining scalability under production constraints.

\vspace{-0.1in}
\section{Method}
% In this section, we present \method (Figure~\ref{fig:framework}), a framework that performs recommendation by reasoning over a heterogeneous social--item interaction graph. Specifically, \method treats recommendation as the discovery of informative reasoning paths that connect a user to relevant items through social and semantic relations. Given the user-centered graph constructed in Section~\ref{sec:graph}, \method selectively discovers paths that reflect social influence, creator affinity, and content relatedness. These paths provide explicit evidence for recommendation, enabling both effective personalization and interpretability. The core challenge is to efficiently discover high-signal paths in a large, noisy graph without enumerating the combinatorial space of all possible paths.

In this section, we present \method, a framework that performs recommendation by reasoning over the heterogeneous social--item graph introduced in Section~\ref{sec:graph}. Rather than aggregating all neighboring information into a single representation, \method treats recommendation as a structured graph exploration problem: starting from a target user, an LLM policy selectively follows typed relations and surfaces candidate items together with explicit evidence paths. The key challenge is to discover high-signal paths in a large and noisy graph without enumerating the combinatorial space of all possible paths.

\begin{figure*}[!htb]
    \centering
    \includegraphics[width=\textwidth]{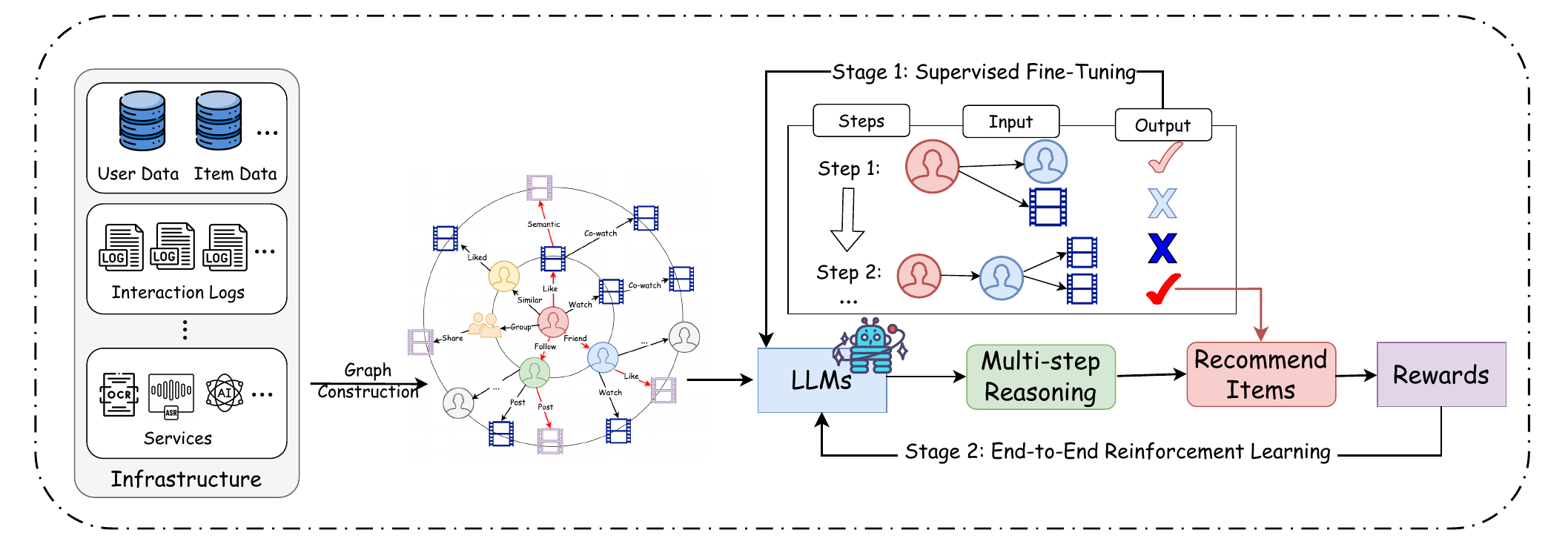}
    \vspace{-0.2in}
    \caption{An overview of the proposed framework \method. We construct a heterogeneous social--item graph from production data and use an LLM policy to perform multi-step graph exploration that surfaces recommended items with evidence paths. The policy is trained in two stages: (1) supervised fine-tuning with step-wise supervision, followed by (2) end-to-end reinforcement learning with rule-based rewards.}
        \vspace{-0.1in}
    \label{fig:framework}
\end{figure*}

\vspace{-0.1in}
\subsection{LLM-guided Graph Exploration}
\label{subsec:explore_rec}

Given a target user $u \in \mathcal{U}$ and a sampled local subgraph $S(u)$, \method incrementally expands partial reasoning paths and surfaces item nodes as recommendations. The LLM policy observes the current path context, the available typed edges, and the semantic descriptions of involved nodes, and then outputs structured exploration actions.

\paragraph{Exploration state.}
We initialize the path set with a length-$0$ path at the target user:
\[
\mathcal{P}_0 = \{p_0\}, \quad p_0 = [u].
\]
At step $d$, the model maintains a set of current paths
\[
\mathcal{P}_d = \{p_1^d,\ldots,p_{n_d}^d\},
\]
where each path ends at a frontier node. The frontier set is
\[
\mathcal{F}_d = \{\mathrm{tail}(p) \mid p \in \mathcal{P}_d\}.
\]
The model observes the outgoing typed neighborhood of these frontier nodes,
\[
\mathcal{N}(\mathcal{F}_d)
=
\bigcup_{v \in \mathcal{F}_d}
\{(v,r,v') \in E(S(u))\},
\]
together with node and relation semantics. This state is serialized into a structured prompt containing the current paths, candidate next-hop edges, and compact textual node profiles.

\paragraph{Exploration action.}
At each step $d$, the policy predicts a structured action with two components:
path expansions and surfaced item--path pairs. The expansion set is
\[
\Delta\mathcal{P}_d \subseteq 
\Big\{
p \oplus (v,r,v') \mid
p\in\mathcal{P}_d,\ 
v=\mathrm{tail}(p),\
(v,r,v')\in\mathcal{N}(\mathcal{F}_d)
\Big\},
\]
where $\oplus$ appends one valid typed hop to the path. The policy may also
surface item nodes reached by selected paths:
\[
\mathcal{E}_d \subseteq
\{(i,p') \mid p'\in\Delta\mathcal{P}_d,\ 
i=\mathrm{tail}(p'),\ i\in\mathcal{I}\}.
\]
Each pair $(i,p')$ represents a recommended item and its evidence path.
The next-step path set is updated as
\[
\mathcal{P}_{d+1}=\Delta\mathcal{P}_d.
\]

\paragraph{Termination and outputs.}
The exploration terminates when $\mathcal{P}_{d+1}=\emptyset$ or when
the maximum depth $D_{\max}$ is reached. The final evidence set and
surfaced item set are
\[
\mathcal{E}=\bigcup_{d=0}^{D_{\max}}\mathcal{E}_d,
\qquad
\mathcal{C}=\{i\mid(i,p)\in\mathcal{E}\}.
\]
The output $\mathcal{C}$ is used for recommendation evaluation, while $\mathcal{E}$ provides explicit paths that explain how the model connects the target user to each surfaced item.

We use an LLM as the exploration policy because the action requires both structured graph reasoning and semantic understanding of heterogeneous node descriptions. However, an off-the-shelf LLM is not directly optimized for relation-consistent graph traversal or recommendation utility. We therefore train the policy in two stages: supervised fine-tuning (SFT) to warm-start valid path generation, followed by reinforcement learning (RL) to optimize full-rollout recommendation quality.

% \paragraph{Termination and outputs.}
% The exploration terminates when $\mathcal{P}_{d+1}=\emptyset$ (no further expansions are selected) or when a maximum depth $D_{\max}$ is reached. The final outputs include (i) the union of all surfaced items,
% \[
% \mathcal{C}=\bigcup_{d=0}^{D_{\max}} \mathcal{C}_d,
% \]
% and (ii) the set of surfaced item--path pairs
% \[
% \mathcal{E}=\bigcup_{d=0}^{D_{\max}}\{(i,p)\mid i\in \mathcal{C}_d,\ p\in \mathcal{P}_d,\ \mathrm{tail}(p)=i\},
% \]
% which provides explicit evidence paths for the recommended items.

% We adopt a large language model (LLM) as the exploration policy due to its strong semantic understanding and multi-step reasoning capability. However, off-the-shelf LLMs are pretrained on general-purpose corpora and do not natively specialize in recommendation objectives or structured graph exploration. To bridge this gap, we employ a two-stage post-training strategy: \textbf{Stage I} performs supervised fine-tuning (SFT) to warm-start the policy for relation-consistent, preference-aligned exploration, and \textbf{Stage II} applies end-to-end reinforcement learning (RL) to directly optimize rollout-level recommendation outcomes.

% \vspace{-0.1in}
\subsection{Stage I: Supervised Fine-Tuning (SFT)}
\label{subsec:sft}

The goal of Stage~I is to warm-start the exploration policy so that it can (i) generate relation-consistent expansions on the heterogeneous graph, and (ii) imitate user engagement behaviors by reaching engaged items through coherent reasoning chains. We construct supervised training trajectories from historical user--item engagements and fine-tune the LLM to imitate these trajectories.

\paragraph{Trajectory construction.}
For each positive engagement $(u,i^+)$, we enumerate valid shortest paths from $u$ to $i^+$ in the sampled subgraph $S(u)$ and treat them as reference reasoning paths. These paths induce step-level state--action pairs
\[
\mathcal{D}_{\mathrm{SFT}}=\{(s_d,a_d^*)\}.
\]
At step $d$, the state $s_d$ contains the current path set $\mathcal{P}_d$, frontier set $\mathcal{F}_d$, and local neighborhood $\mathcal{N}(\mathcal{F}_d)$. The gold action $a_d^*$ is set-valued: it includes all expansions that follow at least one reference shortest path and all item--path pairs that reach a target item at this step. The action is serialized using the same structured output format used during inference.

\paragraph{SFT objective.}
We fine-tune the LLM policy $p_\theta$ by minimizing the token-level negative log-likelihood of the serialized gold actions:
\[
\mathcal{L}_{\mathrm{SFT}}
=
\sum_{(s_d,a_d^*)\in \mathcal{D}_{\mathrm{SFT}}}
-\log p_\theta(a_d^* \mid s_d).
\]
This stage equips the policy with basic graph-navigation ability, relation consistency, and preference-aligned path generation, providing a strong initialization for RL.

% \paragraph{Trajectory construction.}
% We construct the supervised fine-tuning dataset as a collection of step-level state--action pairs,
% $\mathcal{D}_{\mathrm{SFT}}=\{(s_d, a_d^{*})\}$.
% At step $d$, the state $s_d$ summarizes the current exploration context, including the current path set $\mathcal{P}_d$ (and frontier $\mathcal{F}_d$) and the local neighborhood $\mathcal{N}(\mathcal{F}_d)$.
% To derive supervision, for each observed positive user--item engagement $(u,i^+)$ we enumerate \emph{all valid shortest paths} from $u$ to $i^+$ in the sampled subgraph and treat these paths as reference reasoning signals. Aggregating across all shortest paths yields a \emph{set-valued} gold action at each step: $a_d^{*}$ consists of (i) the set of path expansions that extend any current frontier path along at least one shortest path, and (ii) the set of target items that become reachable (i.e., appear as terminal nodes of discovered paths) at this step. Accordingly, the model takes $s_d$ as input and predicts an action $a_d=(\mathcal{C}_d, \mathcal{P}_{d+1})$.

% \paragraph{SFT objective.}
% We fine-tune the LLM by minimizing the token-level negative log-likelihood over state--action pairs:
% \[
% \mathcal{L}_{\mathrm{SFT}}
% = \sum_{(s_d,a_d^{*})\in \mathcal{D}_{\mathrm{SFT}}}
% -\log p_\theta(a_d^{*} \mid s_d),
% \]
% where $p_\theta$ is the LLM policy. This stage equips the policy with basic graph-navigation competence and produces coherent, valid reasoning paths, providing a strong initialization for the subsequent RL stage.

\vspace{-0.1in}
\subsection{Stage II: End-to-End Reinforcement Learning}
\label{subsec:rl}

Although SFT provides step-wise supervision, recommendation quality depends on the entire rollout: a sequence of locally plausible expansions may still surface suboptimal items. We therefore further optimize the policy with reinforcement learning using deterministic, rule-based rewards.

\paragraph{Rollout.}
For each training instance (a target user $u$ together with one sampled local subgraph $S(u)$), the current policy $\pi_\theta$ performs step-by-step exploration as described in Section~\ref{subsec:explore_rec}, producing a sequence of actions $\{a_d\}_{d=0}^{D_{\max}}$, where $a_d=(\mathcal{C}_d, \mathcal{P}_{d+1})$. The rollout yields final outputs $(\mathcal{C},\mathcal{E})$, where $\mathcal{C}=\bigcup_d \mathcal{C}_d$ is the set of surfaced items and $\mathcal{E}$ contains the corresponding item--path evidence traces.

\paragraph{Reward design.}

To enable flexible and interpretable optimization, we adopt a \emph{rule-based} RL framework. Specifically, we design three complementary reward components:

\noindent \textbf{(1) Format reward.}
The policy must output actions in an executable schema, e.g.,
\[
\texttt{\{"surfaced\_items":[...], "expansion\_paths":[...]\}}.
\]
We set $\mathcal{R}_{\mathrm{fmt}}=1$ if all steps are schema-compliant and correspond to valid typed edges in the sampled graph, and $\mathcal{R}_{\mathrm{fmt}}=-1$ otherwise. 
% This reward encourages outputs that can be directly parsed and executed by the graph exploration system.

\noindent \textbf{(2) Final recommendation reward.}
Since each rollout is performed on a sampled subgraph, we evaluate recommendations against the reachable positives within that subgraph. Let $\mathcal{I}^+(u)$ be the set of items that user $u$ engaged with and define
\[
Y_{S(u)}=\mathcal{I}^+(u)\cap \mathcal{I}(S(u)),
\]
where $\mathcal{I}(S(u))$ denotes the item nodes in $S(u)$. Given the final surfaced item set $\mathcal{C}$, we use the $F_1$ score as the rollout-level recommendation reward:
\[
\mathcal{R}_{\mathrm{rec}} = F_1(\mathcal{C},Y_{S(u)}).
\]
This reward balances precision and recall over the items that are reachable from the sampled graph.

\noindent \textbf{(3) Step-wise shaping reward.}
To guide intermediate exploration, we reward expansions that move toward reachable positive items. At step $d$, let $Z_d$ denote the set of frontier-neighbor nodes that lie on at least one shortest path from the current frontier to an item in $Y_{S(u)}$, and let $\hat{Z}_d$ denote the newly reached nodes induced by the predicted expansions. We define
\[
\mathcal{R}_{\mathrm{step}}(\hat{Z}_d,Z_d)=
\begin{cases}
-1.0, & |\hat{Z}_d|=0,\\
-0.9, & |Z_d|=0,\ |\hat{Z}_d|>0,\\
-\eta(|\hat{Z}_d|), & |\hat{Z}_d\cap Z_d|=0,\ |Z_d|>0,\\
F_1(\hat{Z}_d,Z_d), & |\hat{Z}_d\cap Z_d|>0.
\end{cases}
\]
The first case penalizes empty expansions, the second penalizes unnecessary expansion when no reachable positive path exists, and the third penalizes expansions that miss all high-signal nodes. We set $\eta(1)=0.3$ and $\eta(|\hat{Z}_d|\geq 2)=0.5$, assigning
a mild penalty to a single incorrect expansion and a larger penalty to excessive incorrect branching.

\paragraph{Overall reward.}
The final rollout reward is a weighted combination of the three components:
\[
\mathcal{R}
=
\alpha_{\mathrm{fmt}}\mathcal{R}_{\mathrm{fmt}}
+
\alpha_{\mathrm{rec}}\mathcal{R}_{\mathrm{rec}}
+
\alpha_{\mathrm{step}}
\frac{1}{D_{\max}+1}
\sum_{d=0}^{D_{\max}}
\mathcal{R}_{\mathrm{step}}(\hat{Z}_d,Z_d).
\]
We set $\alpha_{\mathrm{rec}}=0.5$, $\alpha_{\mathrm{step}}=0.3$, and $\alpha_{\mathrm{fmt}}=0.2$, prioritizing final recommendation quality while still enforcing valid and selective exploration.
We optimize the policy using Group Relative Policy Optimization (GRPO)~\cite{guo2025deepseek}.

\vspace{-0.1in}
\section{Experiments on Public Datasets}
\begin{table*}[!htb]
\centering
\caption{The performance comparison on Delicious and Foursquare datasets. The best and second-best are in \textbf{bold} and \underline{underline}.}
\label{tab:performance}
\resizebox{\textwidth}{!}{%
\begin{tabular}{c|c|cccc|cccc}
\hline
\multirow{2}{*}{Category} & \multirow{2}{*}{Method} & \multicolumn{4}{c|}{Delicious} & \multicolumn{4}{c}{Foursquare} \\ 
 &  & Recall@5 & Precision@5 & Recall@20 & Precision@20 & Recall@10 & Precision@10 & Recall@50 & Precision@50 \\ \hline
{MF-based} & TrustMF & 0.0063 & 0.0061 & 0.0154 & 0.0053 & 0.0266 & 0.0380 & 0.0703 & 0.0206 \\ \hline
 \multirow{3}{*}{GNN-based} & GraphRec & 0.0028 & 0.0060 & 0.0195 & 0.0122 & 0.0119 & 0.0176 & 0.0366 & 0.0111 \\ 
 & MHCN & 0.0230 & 0.0412 & 0.0753 & 0.0374 & 0.0346 & 0.0500 & 0.0863 & 0.0258 \\
 & SEPT & 0.0218 & 0.0421 & 0.0739 & 0.0376 & 0.0287 & 0.0410 & 0.0734 & 0.0214 \\ \hline
\multirow{2}{*}{Influence Propagation} & DiffNet & 0.0053 & 0.0041 & 0.0107 & 0.0024 & 0.0218 & 0.0317 & 0.0629 & 0.0186 \\
 & DiffNet++ & 0.0094 & 0.0206 & 0.0312 & 0.0182 & 0.0157 & 0.0227 & 0.0478 & 0.0141 \\ \hline
\multirow{2}{*}{Noise-robust} & RecDIFF & 0.0181 & 0.0380 & 0.0659 & 0.0354 & 0.0185 & 0.0276 & 0.0484 & 0.0147 \\
 & SGIL & 0.0193 & 0.0354 & 0.0645 & 0.0304 & 0.0140 & 0.0199 & 0.0560 & 0.0163 \\ \hline
\multirow{3}{*}{LLM-based} & BIGRec (3B) & 0.0039 & 0.0058&0.0112 & 0.0046& 0.0352& 0.0487& 0.1039& 0.0303\\
% & COLLM & & & & & & & & \\
 & \method (3B)& \underline{0.0343} & \underline{0.0787} & \underline{0.0872} & \underline{0.0661} & \underline{0.0818} & \underline{0.1108} & \underline{0.2040} & \underline{0.0586} \\ 
  & \method (8B) & \textbf{0.0631} & \textbf{0.0930} & \textbf{0.1374} & \textbf{0.0841} & \textbf{0.0966} & \textbf{0.1300} & \textbf{0.2084} & \textbf{0.0679} \\ \hline
\end{tabular}%
}
\end{table*}

% In this section, we conduct experiments on public datasets to validate the effectiveness of the proposed \method.

% \vspace{-0.1in}
\subsection{Experimental Settings}

\noindent \textbf{Datasets and Evaluation Metrics.} We evaluate on two representative public datasets, Delicious~\cite{Cantador:RecSys2011} and Foursquare~\cite{yang2019revisiting, yang2020lbsn2vec++}, both of which provide user--item interactions, social relationships, and item-side textual information.
% For the Foursquare dataset, we use a sampled subset consisting of 3{,}000 users and their friends, and convert each item's geographic coordinates (latitude/longitude) into a textual location description. 
We split each dataset into 60\%/20\%/20\% for training, validation, and testing, respectively. We adopt Recall@$K$ and Precision@$K$ as evaluation metrics since \method\ outputs an \emph{unordered} set of items. 

\noindent \textbf{Baselines.} We compare \method\ with representative social recommendation methods from several categories: (i) MF-based social regularization (\textsc{TrustMF}~\cite{yang2016social}); (ii) GNN-based social recommenders (\textsc{GraphRec}~\cite{fan2019graph}, \textsc{MHCN}~\cite{yu2021self}, \textsc{SEPT}~\cite{yu2021socially}); (iii) influence propagation models (\textsc{DiffNet}~\cite{wu2019neural}, \textsc{DiffNet++}~\cite{wu2020diffnet++}); and (iv) robustness-oriented methods for noisy social graphs (\textsc{RecDIFF}~\cite{li2024recdiff}, \textsc{SGIL}~\cite{yang2025invariance}). We also include an LLM-based recommender without social information (\textsc{BIGRec}~\cite{bao2023bigrec}). For \textsc{BIGRec}, we use Llama~3.2-3B-Instruct as the backbone LLM; for \method, we evaluate both Llama~3.2-3B-Instruct and Llama~3.1-8B-Instruct~\cite{dubey2024llama} as backbone policies.

\vspace{-0.1in}
\subsection{Overall Performance Comparison}

The overall performance on the Delicious and Foursquare datasets is reported in Table~\ref{tab:performance}. We make the following observations:

\begin{itemize}[leftmargin=1em]
    \item \method consistently achieves the best performance across all evaluation metrics on both datasets, outperforming all baselines by a clear margin.
    \item GNN-based social recommenders such as \textsc{MHCN} and \textsc{SEPT} generally outperform other baselines, highlighting the importance of explicitly modeling graph structure in social recommendation.
    \item Compared with the LLM-based recommender \textsc{BIGRec}, which does not incorporate social information, \method yields substantially better results. Notably, \textsc{BIGRec} even underperforms several traditional social recommendation methods on Delicious, underscoring that semantic modeling alone is insufficient without social relational context.
    \item Comparing different backbone sizes, \method\ with the 8B LLM consistently outperforms its 3B counterpart, indicating that stronger reasoning capacity in larger LLMs further benefits path-based graph exploration and recommendation.
\end{itemize}

% \vspace{-0.1in}
\subsection{Ablation Study}
We conduct ablation studies to verify the effectiveness of the proposed two-stage training strategy. Specifically, we compare three variants: (i) the original \textsc{Llama~3.2-3B-Instruct} without any task-specific adaptation, (ii) \method\ with only supervised fine-tuning (\method\ SFT-only), and (iii) the full \method. Specifically, we report Recall@5 on the Delicious dataset and Recall@10 on the Foursquare dataset. 

\begin{figure}[!htb]
    \centering
    \includegraphics[width=0.8\linewidth]{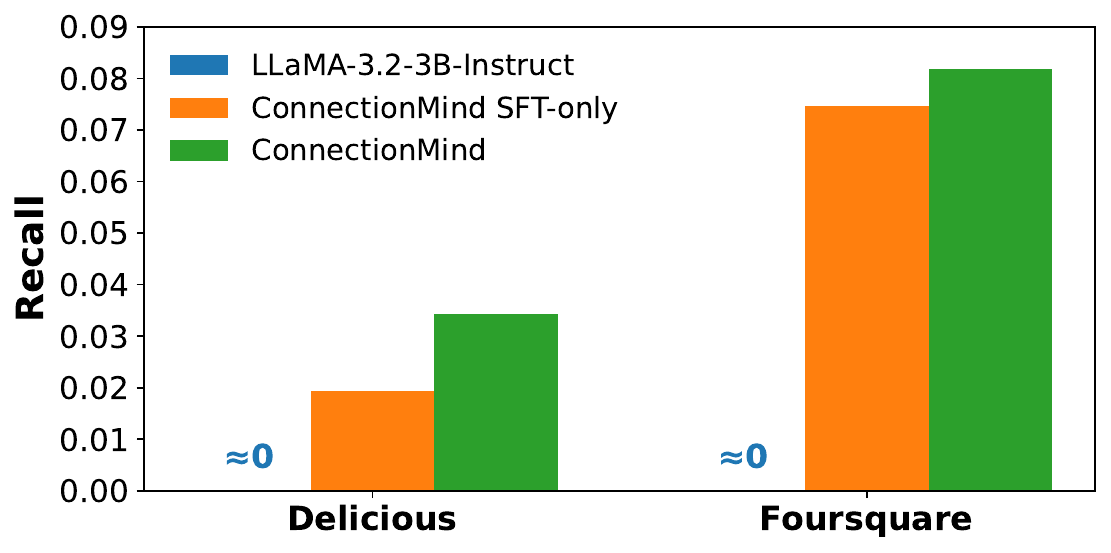}
    \vspace{-0.1in}
    \caption{Performance comparison of different variants.}
    \vspace{-0.1in}
    \label{fig:ablation}
\end{figure}

Overall, Figure~\ref{fig:ablation} shows that task-specific post-training is essential. The original \textsc{Llama~3.2-3B-Instruct} achieves near-zero recall on both datasets, indicating that a pretrained LLM without adaptation cannot perform recommendation or graph exploration. This is because the model must learn both (i) how to follow the structured exploration interface and (ii) user-specific engagement patterns. In contrast, \method\ SFT-only substantially improves performance by learning graph exploration and preference signals from logged supervision, and the full \method\ further boosts recall with end-to-end RL, validating the benefit of optimizing rollout-level exploration and item surfacing.

\vspace{-0.1in}
\section{System Deployment and Evaluation}
\label{sec:online}
Deploying \method at the scale of Meta's short-form video platform, serving billions of users, requires a rigorous balance between the high computational cost of generative reasoning and the strict latency constraints of real-time recommendation. In this section, we detail the production infrastructure, our hybrid inference strategy, and the empirical results from both offline evaluation and large-scale online A/B test.

\vspace{-0.1in}
\subsection{Hybrid Inference Infrastructure}
\begin{figure}[!htb]
    \centering
    \includegraphics[width=0.9\linewidth]{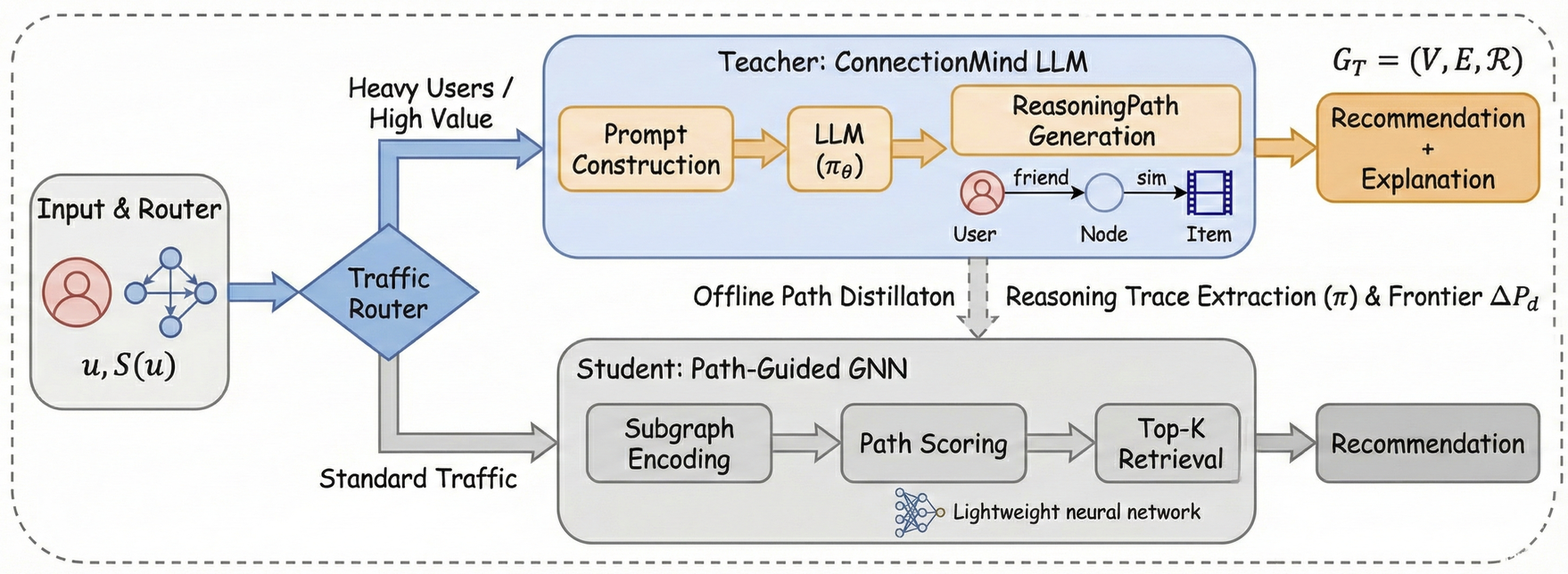}
    \caption{The Hybrid Inference Architecture. The system routes heavy users to the ConnectionMind for explicit reasoning path generation ($\pi_\theta$), while standard traffic is served by a lightweight Student GNN. The Student is trained via offline distillation to mimic the teacher's reasoning traces ($\pi$) without the inference overhead}
    \label{fig:hybrid_inference}
\end{figure}

To address the latency bottleneck of autoregressive generation, we devised a \textbf{Teacher-Student Hybrid Inference} strategy that segments traffic based on user activity profiles (Figure~\ref{fig:hybrid_inference}):
\begin{itemize}[leftmargin=1em]
    \item \textbf{Direct LLM Reasoning (Heavy Users):} The top 5-10\% of most active users exhibit complex social graphs and high engagement velocity. For this cohort, requests are routed to the full ConnectionMind LLM. The high computational cost is justified by the significant value and retention leverage of this user group.
    \item \textbf{Path-Guided Distillation (Standard Users):} For the remaining majority of traffic, we employ an offline distillation pipeline. The LLM processes historical contexts to discover effective meta-paths. These paths are extracted to train a lightweight Graph Neural Network (Student GNN). The student model learns to emulate the teacher's path-finding logic but executes with millisecond-level latency, scaling the benefits of reasoning-aware recommendation to the entire population.
\end{itemize}

\vspace{-0.1in}
\subsection{Experimental Results}
We evaluate the system using both offline historical logs and live production traffic.

\paragraph{Offline Evaluation}
We compare ConnectionMind with Llama 3.1-8B-Instruct backbone against a production \textbf{GNN Baseline} and a standard \textbf{Llama 3.3-70B-Instruct} (ranking without graph reasoning). Using \textbf{Recall@10} as the primary metric, our approach achieves an \textbf{88\% relative lift} over the production baseline (Table~\ref{tab:results}). Notably, the graph-reasoning integration outperforms the vanilla LLM by a wide margin, confirming that semantic knowledge alone is insufficient without structural reasoning.

\begin{table}[h]
    \centering
    \caption{Offline performance (Recall@10) and online A/B test results (relative lift over the production system).}
    \label{tab:results}
    \resizebox{\linewidth}{!}{
    \begin{tabular}{l|c|ccc}
        \toprule
        \multirow{2}{*}{\textbf{Model}} 
        & \textbf{Offline} 
        & \multicolumn{3}{c}{\textbf{Online Metrics (A/B Test)}} \\
        & \textit{Recall@10} 
        & \textit{Exposure} 
        & \textit{Watch Time} 
        & \textit{Video Sessions} \\
        \midrule
        Production GNN 
        & Ref. (100\%) 
        & -- 
        & -- 
        & -- \\
        Llama~3.3-70B-Instruct 
        & +39\% 
        & -- 
        & -- 
        & -- \\
        \textbf{ConnectionMind} 
        & \textbf{+88\%} 
        & \textbf{+0.33\% ± 0.08\%} 
        & \textbf{+0.43\% ± 0.14\%} 
        & \textbf{+0.22\%  ± 0.13\%} \\
        \bottomrule
    \end{tabular}
    }
\end{table}

\paragraph{Online A/B Test}
We conducted an A/B test on a massive user population (tens of millions of users) over a continuous multi-week period, ensuring robustness. The results (Table~\ref{tab:results}) highlight three key business impacts:
\begin{enumerate}
    \item \textbf{Catalog Diversity (+0.33\% Exposure):} The semantic graph densification successfully surfaced long-tail content to cold-start users, proving that reasoning capabilities can break popularity feedback loops.
    \item \textbf{Engagement Depth (+0.43\% Watch-time):} By providing socially grounded recommendations (e.g., explaining \textit{why} a video is relevant via social connections), the model increased user trust and consumption depth.
    \item \textbf{Retention (+0.22\% Video-sessions):} The lift in daily Video-sessions suggests that socially relevant recommendations foster a stronger sense of community, increasing the frequency of app opens.
\end{enumerate}

% While percentage gains in a mature, billion-user optimization environment are typically incremental, these lifts are statistically significant and translate to substantial improvements in total platform time spent.

\paragraph{Summary.}
Overall, our production deployment demonstrates that \method\ can be integrated into a latency-critical recommendation stack via a teacher--student hybrid inference design. Offline, ConnectionMind achieves a substantial Recall@10 improvement over both the production GNN baseline and a vanilla LLM ranker, and online A/B test shows consistent, statistically significant lifts across Exposure, Watch Time, and Video Sessions. While percentage gains in a mature, billion-user optimization environment are typically incremental, these improvements translate to substantial increases in total platform time spent at scale.

\section{Conclusion}
In this paper, we present \method, a reasoning-driven framework for social recommendation deployed at Meta, which formulates recommendation as multi-step path exploration over a heterogeneous social–item interaction graph. Rather than compressing social context into dense embeddings as in prior social recommenders, \method explicitly exposes evidence paths connecting users to recommended items, enabling selective utilization of social signals and improved interpretability. To equip large language models with structured exploration capabilities and user-specific preference modeling, we adopt a two-stage post-training pipeline consisting of supervised fine-tuning followed by end-to-end reinforcement learning with rule-based rewards. Extensive experiments on public benchmarks, together with results from a large-scale production deployment, demonstrate consistent performance gains, validating the effectiveness of reasoning-based graph exploration for social recommendation under real-world constraints.

% \begin{acks}
% \end{acks}

\section*{GenAI Usage Disclosure}
The authors used generative AI tools only for language polishing, grammar checking, and editing of author-written text. The authors reviewed and revised all edited text and take full responsibility for the final content. No generative AI tool was used to generate experimental results, datasets, or evaluation claims.

\bibliographystyle{ACM-Reference-Format}
\bibliography{mybib}

% \appendix

\end{document}